\documentstyle[11pt,dunk2001_asp,twoside,psfig]{article}
\begin{document}
\title{HI and dark matter in spiral galaxies}
\author{Albert Bosma}
\affil{Observatoire de Marseille, 2 Place Le Verrier,
13248 Marseille C\'edex 4, France}

\begin{abstract}
We still do not have a ``standard" model
for the mass distribution of spiral galaxies. I review various 
methods used to delimit the range of values of the mass-to-light
ratio of the disk, such as spiral structure criteria, the influence
of bars, warped and flaring HI layers, velocity dispersions, and 
features in the rotation curve occuring just beyond the optical
radius. It is not yet clear whether disks are maximal or not. 
\end{abstract}

\section{Ken Freeman's early contribution to the dark matter problem}

In his classic paper on exponential disks (Freeman, 1970),
Ken presents a unifying scheme to describe the photometric properties
of disk galaxies, using a simple law which describes 
the radial distribution of the surface brightness of the disk of a
spiral or S0
galaxy reasonably well. He then works out the dynamics of such disks,
and discusses issues related to disk formation. In the Appendix of
that paper he concentrates on four galaxies, and points out that for two 
of them the available data hint at a discrepancy between the
photometric scalelength of the disk and the expected peak of the
rotation curve if the mass-to-light ratio of the disk is independent
of radius. While for the LMC and the SMC there is no such
discrepancy, the available data for M33 and NGC 300 do indicate one:

{``For two of four systems with weak spheroidal components, the
LMC and the SMC, the turnover radii of the rotation curves predicted
from the photometrically derived gradients $\alpha$ appear to be
consistent with those observed. For NGC 300 and M33, the 21-cm data
give turnover points near the photometric edges of these systems.
These data have relatively low spatial resolution; it they are
correct, then there must be in these galaxies additional matter which
is undetected, either optically or at 21 cm. Its mass must be at least
as large as the mass of the detected galaxy, and its distribution must
be quite different from the exponential distribution which holds for
the optical galaxy.''}

These remarks have been quoted by several authors retracing
the history of the dark matter problem in spiral galaxies (e.g.
Van Albada et al. 1985, Sciama, 1993). It is actually
interesting to examine how the interpretation of the data of these two 
galaxies changed as the instrumentation improved. This is of
intellectual relevance to see how scientific debate advances 
with time : it is
sometimes said that the dark matter problem constitutes a
``revolution'', in the sense of Kuhn's book (Kuhn 1962), or worse
a ``paradigm shift'' -- but then ``what is a paradigm ?'' (cf.
Lakatos \& Musgrave 1970) --
so what is the contribution of Ken's paper to this revolution ?

Early data on M33 are very confusing, and the most up-to-date 
information available in 1969 is summarized in an IAU Symposium
published in Russian (cf. Demoulin et al. 1969). Dramatic
improvements in the kinematical data on M33 were obtained in the early
70's, first through the work of the Cambridge HI group (Warner et al.
1973), and then by Rogstad et al. (1976). In the latter paper, one of
the key characteristics of the HI distribution was pointed out : the
HI layer is strongly warped beyond the edge of the optical disk. It 
turns out that if the points on the rotation curve in the warped
part are ignored, as done by Kalnajs (1983), one can fit the rotation
curve assuming the mass follows the light. If the
rotation curve is extended with the HI points from the warped parts,
this cannot be achieved anymore (Athanassoula et al. 1987, fig.~5).

For NGC 300, the early HI data are from a Parkes single dish study,
(Shobbrook \& Robinson 1967), and the surface photometry from De Vaucouleurs
\& Page (1962). Ken pointed out the discrepancy between the expected
and observed turnover point of the rotation curve. HI
data with much better angular resolution, analysed by Rogstad et al.
(1979), show that the HI layer is strongly warped beyond the optical
disk. Further study of this galaxy was done by a Puche, Carignan \&
Bosma (1990)
-- a student of a student of Ken, a student of Ken and a 
postdoc of Ken --
which showed that there is indeed a mass discrepancy 
starting well within the optical radius, vindicating Ken's 
1970 remark.

Thus we can say that Ken's study, as he reported it in the Appendix 
of his 1970 paper, pointed for the first time in an elegant manner
to a mass discrepancy
in a galaxy, based on a then novel general description - which is
still standing - of how the luminosity is
distributed in the disk (the exponential law), the idea that the
mass-to-light ratio in a disk could well be constant, and appealing 
to HI studies to get information on galaxy rotation curves.

Dark halos came later (Ostriker \& Peebles 1973), and their
implication that spiral galaxy masses could really be large a year
after (Einasto et al. 1974, Ostriker et al. 1974). These studies based
themselves on the early work by Roberts \& Rots (1973) and Rogstad \&
Shostak (1972).
Yet at the
observational side a long debate was spun about the validity of
the data (in particular for M31, e.g. Roberts 1975, Baldwin 1975),
the question whether fits with constant M/L were still possible
(Emerson \& Baldwin 1973, Baldwin 1975),
the technical questions of sidelobes and beam smearing
(e.g. discussions by Sancisi and Bosma following Salpeter
1977),
and the
interpretation of rotation curves derived from tilted ring models
to describe warps (cf. discussion after Toomre 1977). 
Many of these problems were settled using high
resolution HI data from the Westerbork telescope, accompanied by
surface photometry and H$\alpha$ kinematics (Bosma 1978, 1981, Bosma
\& Van der Kruit 1979, Van der Kruit \& Bosma 1978).

The work of Rubin et al. (1978), which later led to a series
of papers on the rotation curves of Sc, Sb and Sa galaxies (Rubin
et al. 1980, 1982, and 1985) was contemporary, but the HI data extends
further out. Moreover, when, in the
early 80s, everybody thought that dark halos are necessary to explain
the flatness of rotation curves, Kalnajs (1983) showed
that for optical rotation data, in four cases, there is no need for
a dark halo, a conclusion corroborated for many more optical rotation
curves by Kent (1986, 1987), Athanassoula et al. (1987), Buchhorn (1992)
as reported in Freeman (1992), Moriondo et al. (1999), and 
Palunas \& Williams (2000).

\section{Current problems with mass models of spiral galaxies}

\subsection{``Standard'' composite mass models}

The current practice is to construct mass models using components :
the bulge and disk components are derived from surface photometry, and
a gas component from the radially averaged HI distribution. For the bulge 
and the disk it is assumed that each have a constant mass-to-light ratio,
independent of radius. The molecular gas is implicitly assumed to
follow the optical light. All components are quadratically added and then
compared with the observed rotation curve. If there is a discrepancy,
which is usually the case in the outer parts for extended HI rotation
curves, a dark halo component is added.

Current debate centers on whether the disk is maximum,
i.e. whether the value of the mass-to-light ratio of the disk is such
that there is a maximum of mass in the disk, and thus a minimum amount
of mass in the dark halo. As already mentioned above, for many optical
rotation curves the radial extent is not large enough to show the
presence of a dark halo, assuming that the disk is maximal. On the
other hand,
for many of the HI rotation curves the data come from tilted ring
models (e.g. Rogstad et al. 1974) fitted to the warped part of the HI
disk. Since warps are not really understood, there is always a
lingering worry that such curves contain small systematics errors
due to the neglect of systematic perpendicular motions.

\subsection{Examination of the underlying hypotheses}

Modern surface photometry (e.g. de Jong 1996) shows clearly that, if
there is any colour gradient in disk galaxies, it is in the sense that
disks become bluer in the outer parts.
This means that the mass-to-light ratio in the
disk could be decreasing with radius, which leads, in the composite
models, to
an increase in the concentration index of the dark halo. It seems best
to use the K-band to get the stellar disk mass distribution, but 
there are as yet still few galaxies for which there are
both the requisite K-band photometry and detailed HI and H$\alpha$ 
kinematics. Since the molecular gas scales better with the B-band
(Young \& Scoville 1982), a minor error can be
introduced on this account.

A comparison between the HI rotation data by Verheijen (1997) for Ursa
Major galaxies and my old collection of rotation curves (Bosma 1978)
shows that, while it easy to increase the number of galaxies for which
there is HI data, the quality of the curves does not increase
dramatically. This is due to the fact that it takes quite a lot of
telescope time to get really good HI data, and for the more distant
galaxies this is simply not available.

Beam smearing is another problem, which, while thoroughly understood
in principle,
is still playing an unnecessary large
role in the discussion. Partly this is due
to the unfortunate circumstance that for late type spirals the 
differences between the H$\alpha$ and HI rotation curves are not very
large, and sometimes so small within the admittedly large errors
of both datasets, that some people are misled into assuming 
that they can be ignored. 
This is sometimes justified {\it a posteriori} by taking a high
resolution H$\alpha$ spectrum along the major axis, which agrees
with the HI rotation curve
in some cases, but not in others. Systematic programs are now underway
to remedy this situation, using several techniques : long slit
spectroscopy of emission lines such as H$\alpha$ (e.g. de Blok et al.
2001), 3D imaging using Fabry-P\'erot techniques (e.g. Corradi et al. 1991,
Blais-Ouellette et al. 2001), or CO data
from e.g. the BIMA interferometer (Wong 2000).

High resolution is crucial in the debate about the predicted slopes 
for the density profiles
of the dark matter in LSB galaxies -- inner power law slopes of order 
-1.5 (Moore et al. 1999, Fukushige \& Makino 2001) or 
-1.0 (Navarro et al. 1996) -- and those observed,
which are of order 0.0, as argued e.g.
in De Blok et al. (2001) and De Blok \& Bosma (2001).

\subsection{Is the dark matter directly associated with the HI ?}

A relation between the HI gas mass density and the total mass density
has been pointed out by Bosma (1978, 1981), who found that the ratio
of gas mass surface density to total mass surface density becomes
constant in the outer parts. This implies that the HI and the
dark matter are somehow related, and inspired Pfenniger \& Combes
(1994) to postulate that the dark matter is in cold fractal molecular gas. 
In the context of composite mass models, the
HI gas rotation curve is similar to the one of the dark halo, at least
for maximum disk models, as discussed by Carignan et al. (1990).
This is of course in some sense a restatement of the
original result. Note that in this approach any molecular gas as inferred
from the CO-distribution is
assumed to scale as the light distribution.

Hoekstra et al. (2001) examine this further for a sample of well
observed spiral galaxies already discussed in Broeils (1992). They 
calculate for 24 galaxies a histogram of the ratio 
$\Sigma_{\rm dark}$/$\Sigma_{\rm gas}$, determined by scaling up the
HI rotation curve so that the observed rotation curve is fitted by 
a combination of disk, bulge, and scaled gas. The average ratio of
$\Sigma_{\rm dark}$/$\Sigma_{\rm gas}$ is about 7, but with some
spread. If one includes in this histogram the values given for dwarf
galaxies determined by Swaters (1999), the distribution becomes quite 
dispersed, with no single peak.

\section{Are galactic disks self-gravitating ?}

In this section I will summarize the attempts to pin down more
precisely the value of the mass-to-light ratio of the disk, in order
to ascertain whether disks are maximal or not. A convenient way to 
put a number to this question is to consider 
a parameter $\gamma$, which is the ratio of the maximum amplitude 
of the disk rotation curve to that of the total rotation curve, evaluated
at 2.2 disk scalelengths (which is roughly the radius of the peak of 
the disk rotation curve)~: 

\begin{equation}
\gamma = \biggl({V_{disk} \over V_{total}}\biggr)_{~at~R~=~2.2~h}
\end{equation}

Note that the maximum value $\gamma$ can attain depends on whether there 
is a bulge present in the galaxy. If a bulge is present, $\gamma$ can 
not reach unity even in the case of ``maximum disk''. In what follows,
I will rapidly review various methods to pin down $\gamma$, 
as an update to the discussion in Bosma (1999).

\subsection{constraints from spiral structure}

\subsubsection{Swing amplifier criteria :}

Athanassoula et al. (1987) have discussed the application of spiral 
structure constraints to composite mass models. They examine for each
model whether there is the possibility of swing amplification based
on the mechanism of spiral amplification discussed by Toomre (1981).
A more graphical description is given in Bosma (1999). The physics of 
the swing amplifier depend on the shape of the rotation curve and on
a characteristic X parameter, which in turn depends on the epicyclic 
frequency $\kappa$, the number of arms m, and the active surface mass 
density of the disk.

The range in mass-to-light ratio varies with about a factor of two if 
one requires swing amplification of the m = 2 perturbations : the
lower limit is set by requiring that the disk is massive enough so
that amplification of the m = 2 perturbations is just allowed, and an
upper limit is set by requiring that
amplification of the m = 1 perturbations is just prohibited. Usually the
latter condition is fulfilled if one requires a model with maximum
disk and a halo with non-hollow core. See Athanassoula et al. (1987)
for more details.

\subsubsection{Velocity signatures of spiral arm perturbations :}

Already in the M81 HI data obtained with the Westerbork telescope
(Rots \& Shane 1974) the effects of peculiar motions due to the spiral
arms are clearly seen. These were modeled with a response calculation
by Visser (1980),who did not use a dark halo in his models.
Lately, Alfaro et al. (2001) show a single long-slit
spectrum of the galaxy NGC 5427, where the presence of ``wiggles'' in
the position-velocity curve are quite clearly associated with the
spiral arms. It is clear that the presence of such wiggles indicates
that the disk is self-gravitating enough to produce them.

In what promises to be the first of a series of papers, Kranz et al. 
(2001) present long-slit data for NGC 4254, a spiral galaxy in the 
Virgo cluster for which also HI data are available from Phookun et al.
(1993). They try to reproduce the observed velocity perturbations with
a stationary gas flow model using the K-band image of the
galaxy as input to the evalution of the disk part of the galactic 
potential. They find that a maximum disk model produces too
large velocity perturbations, and put an upper limit on $\gamma$ of 0.8.
However, this galaxy is lopsided in the HI, the spiral may be evolving,
and the small bar in the center of the galaxy might have a different
pattern speed than the main spiral pattern.
Moreover, the inclination may not be as low as the authors take it.

\subsection{bar slow down due to dynamical friction}

Debattista \& Sellwood (1998, 2000) performed N-body simulations 
to study the slow down of a bar due to the dynamical friction with
the surrounding dark halo. They argue that realistic bars can be 
made only if their disks are close to maximum.  
Athanassoula (this volume) addresses this subject anew. She finds
that all strongly barred galaxies have maximum disks. Indeed, even if
they start as sub-maximum, the rearrangement of the disk material
due to the formation of the bar will lead to a maximum disk
(cf. also Athanassoula \& Misiriotis 2002).

\subsection{fits of gas flow models to barred spirals}

Weiner et al. (2000) modeled in detail the gas flow in the barred
spiral NGC 4123, 
and find that the best fit to the velocity data
requires a maximum disk model for the mass distribution. Note that 
here again the modelling is done as a response calculation for a 
stationary flow in the potential derived from an optical image. As
in the case of NGC 4254, no time evolution has been considered.

\subsection{warped and flaring HI disks}

Data on the vertical axis ratio of the halo, c/a, have been collected
and discussed by Olling and Merrifield (2000). 
Several methods are used, but for spirals the
favourite method is that of the flaring of the HI layer (cf. Olling 1996).
Relatively few galaxies have been studied this way, and
more work on this is in progress (collaboration OBrien, Bosma \& Freeman).

Warps also could in principle tell us something about the dark
halo. If they are steady, and the halo is misaligned with the disk,
MOND could in principle be ruled out. However, warps may well evolve
rather rapidly, as simulations attest (Kuijken \& Dubinski 1995). It
is then not clear whether the presently observed HI warps can be
used as a diagnostic of halo properties. Bosma (1991) showed that 
the frequency of warps is rather high, a result confirmed by more
recent data of the WHISP survey. Bosma (1991) also
discussed a relation between warps and halo properties :
warps avoid the region where the ratio of halo core radius to optical
radius is small. More work on this is needed.

\subsection{stellar velocity dispersions in disks}

Bosma (1999) and Fuchs (1999) report efforts to use stellar
velocity dispersions to pin down $\gamma$, using data from Bottema
(1993), or newer data of comparable quality. The series of assumptions
made by Bottema (1993) were abandoned in those works, and the
conclusion is that, strictly speaking, maximum disk models lead
to Q-values slightly less than 1. However, the constant 3.36 in the
formula for the critical velocity dispersion is dependent on the
assumptions for the shape of the velocity ellipsoids, on the thickness
of the layer, and on the presence of gas. Hence there is quite a bit
of intrinsic uncertainty to this method. Moreover, a recent paper
by Vega Beltran et al. (2001) shows that at the observational side
the error bars are rather large. Nevertheless, further work on this
is warranted, given the fundamental importance of velocity dispersions
for the dynamics of spiral galaxies.

\subsection{attempt to refute a MOND prediction}

For a number of well observed galaxies, Begeman et al. (1991) have
worked out maximum disk models as well as mass models based
on MOdified Newtonian Dynamics (Milgrom 1983). They
conclude that MOND fits the data equally well, and determine
the universal constant in this theory, the value of the critical acceleration
parameter a$_0$, to be 1.21 10$^{-8}$ cm$^{-2}$, 
by averaging the best fit values for a number of spirals.
However, to fit the galaxy 
NGC 2841 with this value, it is required that its distance is  
about twice the
distance inferred from the Hubble flow, i.e. 19.5 Mpc instead of 9.5
Mpc.

A recent study by Macri et al. (2001) determines the distance to this
galaxy using Cepheids discovered on HST images. They derive a final
value of 14.1 $\pm$ 1.5 Mpc. In collaboration with Erwin de Blok, I
looked again at the data : we took some H$\alpha$ data in the inner
parts, and tried to fit a mass model based on MOND. The model is
uncertain in the inner parts; in the
outer parts, its prediction is below the observed curve. But here
again we are in the warped part of galaxy, hence we cannot yet rule 
out MOND with certainty using these data.

\section{Extent of rotation curves}

HI studies are limited to radii of about 30 - 50 kpc for a large
galaxy, and rarely reach beyond that. The deepest HI image of a spiral
galaxy is the one obtained by Van Gorkom et al. (unpublished), for the
galaxy NGC 3198. As discussed in Maloney (1993), the HI in the outer
parts drops relatively suddenly, but the total gas density may not
drop as abruptly. Efforts to detect the ionized gas
supposedly surrounding the HI at large radii have remained unsuccesful.

Further information at larger radii can be obtained by the study of
satellites. Since most galaxies do not have many companions, a
statistical treatment is in order. Zaritsky \& White (1994) 
and Zaritsky et al. (1997) have done this for a large sample of
spiral galaxies and conclude that dark halos are indeed extended.
Weak lensing studies
using data from the Sloan Digital Sky Survey (McKay et al. 2001)
confirm these results and show that
the size of a typical halo is about 260 h$^{-1}$ kpc, which amounts,
for a Hubble constant of 70 km s$^{-1}$ Mpc$^{-1}$, to 364 kpc --
half the distance between our Galaxy and M31.

It may well be asked whether the rotation curve of such an extended
halo stays flat, or whether a decline sets in. For our Galaxy, this
question has been addressed in several papers (e.g.
Zaritsky et al. 1989), and the result seem to be depend on 
the exact radial velocity of the Leo II system. 
For other galaxies,
from HI data alone, the study of Verheijen (1997, 2001) shows 
for bright galaxies
that there is a slight difference between V$_{\rm max}$ and V$_{\rm flat}$,
the latter referring to the rotation speed of the HI envelope. The 
Tully-Fisher relation tightens significantly
when one uses the relation between the absolute magnitude in the
K-band and V$_{\rm flat}$ instead of V$_{\rm max}$. This comes on 
top of the tightening of the relation at the faint end by 
considering a ``baryonic correction'' (cf.
Freeman 1998, McGaugh et al. 1999). Thus the Tully-Fisher relation is 
between the amount of baryons in a spiral galaxy and the rotation speed
of its dark halo.

The difference between V$_{\rm max}$ and V$_{\rm flat}$ is not new,
and for maximum disks it leads to the breakdown of the conspiracy
between disk and halo (cf. Casertano \&
Van Gorkom 1991). Most often, the rotation curve drops rather suddenly
just beyond the optical image, as already noted for NGC's 5033 and 5055
by Bosma (1978, 1981). This feature could by itself be used to argue 
that disks are not far from maximum. However, an attempt to quantify this
properly, for the galaxy NGC 4414 by Bosma (1998), shows that while a
maximum disk model is preferred, a no m = 2 model cannot be excluded 
clearly. From statistics, both for a sample of galaxies
with large rotation speeds I am working on, and one
from the WHISP survey, I find
that roughly 10 - 20 \% of large spirals (Sa's - Sc's) have such 
a feature. This should be quantified further. 

\section{Conclusion}

A comparison with Bosma (1999) and Sellwood (1999)
shows where progress has been made on
the issue of whether disks are self-gravitating:
the results from Athanassoula et al. (1987) based on the swing 
amplifier criteria still stand, new results from Kranz et
al. (2001) favour less than maximum disks, several studies
argue that barred galaxies have maximum disks, and the use of 
stellar velocity dispersions is
hampered by lack of good data. So not much
progress ? Better data will come from the
gravitational lensing methods, which rather soon might come
up with a clear answer. Until then, we can still try to improve
things by getting better data and simulations along the lines
discussed above.

\acknowledgments
Thanks are due to Lia Athanassoula for frequent discussions, and to
Jerry Sellwood for recent exchanges of our points of view. Ken
Freeman gave me freely his advice when I was a student in Groningen
in the seventies (... better be a good observer than a lousy theoritician
...), arranged for a Research Fellowship for me in Stromlo after my
thesis, and kept in contact over the years. In the nineties, we took
up old collaborative work, and started new projects, some of which are
still ongoing. I benefitted a lot from his comments on many topics.
It is a pleasure to acknowledge also the kind hospitality Ken and
Margaret have shown me on numerous occasions.

\newpage
\section*{Discussion}

\medskip
\noindent {\it Van der Kruit:\, } {I would comment that to avoid
confusion we should refrain from redefining the constant in the Toomre
Q parameter, but rather require it to be at least some other value than
1. My question is the following. You ask : ``are galaxy disks
self-gravitating'', and answer probably yes for larger disks. What
do you think the answer is for the Galactic disk in the solar
neighbourhood ?} 

\medskip
\noindent {\it Bosma:\, } {OK for the comment. As for the question, 
the microlensing results in particular indicate that there is a lot of baryonic
matter in the central parts of the Galaxy, so that there does not seem
much room for a cuspy halo, as seemingly required by current models
based on cosmological numerical simulations of cold dark matter, cf.
a recent paper by Binney \& Evans (astro-ph/0108505).}

\medskip
\noindent{\it added after the meeting:\, } {The solar radius is roughly at
2.2 times the scalelength of the disk, and the ``measured'' disk surface
density of 48 M$_{\odot}$ pc$^{-2}$ gives a peak speed of 126 km s$^{-1}$.
It thus really matters whether the bulge/bar system is considered as
part of the {\bf {disk}}, and not as part of a spherical (and thus
stabilizing) {\bf {bulge}} component. This brings the circular
speed of the combined bulge/disk to about 160 km/s at least. Thus $\gamma$
could be of order 0.8 in our Galaxy, or even higher since the
contribution of the molecular gas is quite uncertain.}

\medskip
\noindent {\it King:\, } {As a spectator in this field rather than a
participant, I get the overall impression that because of a lot of
uncertainties (warps, asymmetries, etc., etc.) the estimation of the
characteristics of dark halos is still very uncertain. Is this fair ?}

\medskip
\noindent {\it Bosma:\, } {Yes, this is fair enough. I think that most
people agree about the existence of the mass discrepancies, as brought
out by the HI observations of extended gas disks, and many would say
that implicates dark matter, but other than that 
there is not much agreement on how important the contribution of the
dark matter is in the inner parts of disk galaxies.} 

\medskip
\noindent {\it King:\, } {You would then emphasize the qualitative
conclusions, but not the quantitative ones.} 

\medskip
\noindent {\it Sellwood:\, } {A comment on the Kranz et al. paper :
They assume a constant amplitude, constant pattern speed spiral, which will
induce large amplitude responses at strong resonances. A more realistic
transient spiral pattern may require more mass in the disk to produce the 
observed non-circular motions.} 

\medskip
\noindent {\it Bosma:\, } {As I said in the talk, the authors
acknowledge that some of their assumptions are problematic, including
the question of the stationarity of the spiral pattern for which they 
compute the gas response. Your suggestion may well be right. For this
particular galaxy, NGC 4254, I am myself worried about the inclination
they assume. But I understand that they have more data on other
spirals, so it will be interesting to see what they come up with.}

\end{document}